\documentclass[amsmath,showpacs,amssymb,aps,prb,twocolumn,superscriptaddress,10pt]{revtex4-1}
\usepackage{amsmath}
\usepackage{graphicx}
\usepackage{bm}
\usepackage{xcolor}

\newcommand{\bk}{{{\bf{k}}}}

\begin{document}

\title{Individual band with higher Chern numbers in double perovskite \{001\} monolayers}

\author{A. M. Cook}
 \email{cooka@physics.utoronto.ca}
\affiliation{Department of Physics, University of Toronto,
Ontario, Canada, M5S 1A7.}

\begin{abstract}
We show a tight-binding model for double perovskite monolayers grown in the \{001\} crystallographic direction can exhibit higher Chern numbers for the lowest band of the model dispersion in the absence of an applied magnetic field, including $\pm2$, $\pm4$, $\pm6$, and $\pm8$, for model parameters and terms relevant to Sr$_2$CrWO$_6$ (SCWO).  We further show it is possible to tune between these different topological phases by applying a weak in-plane magnetic field.  These results may be important for experimentally realizing exotic band topologies -- and even quantum anomalous Hall insulators with higher Chern numbers -- based on this physics.
\end{abstract}


\maketitle

\section{Introduction}

The possibility of realizing dissipationless edge states of the quantum Hall effect (QHE) in the absence of an applied magnetic 
field has drawn considerable interest from the physics community for decades, with the first proposal for such a system made by Haldane\cite{haldane000}. 
As a result of more recent proposals\cite{zhang000,ohgushi000,martin000,
nandkishore000,xu000} to realize this physics through a combination of spin-orbit coupling (SOC) and net magnetization or non-coplanar magnetic order, this interest has intensified. Experiments on (Bi,Sb)$_2$Te$_3$ topological insulator (TI) films doped with magnetic Cr atoms have reported the first observation of the QAH effect\cite{chang000} and, more recently, robust QAH states have been realized in hard ferromagnetic topological insulators\cite{chang001}. 

It is important to note, however, that the above proposals realize systems with Chern number $\pm1$, emerging from Chern number $\pm1$ band topology. Systems with higher Chern numbers and higher Chern number topology for individual bands are also of great interest for practical and fundamental reasons. For instance, models for Chern insulators with Chern number 2 originating from a single filled band have been shown to exhibit a novel nematic QAH phase\cite{cook001}. Fractional filling of Chern insulators with Chern number $C=2$ is also anticipated to lead to new topological states with novel elementary excitations\cite{barkeshli000}. The dissipationless edge modes of the QAH insulator have furthermore been proposed as interconnects for integrated circuits\cite{zhang003}, and a QAH insulator with higher Chern number lowers the contact resistance expected at contact points between QAH insulator components and the rest of such circuits, reducing losses\cite{sczhang000}. 

Although systems exhibiting Chern numbers larger than $1$ exist\cite{zhang001,jiang000,wright000,qiao000,tse000,jzhang000,chen000,ding000,shiba000,cook000},  most of these realize a total Chern number of $\pm2$ at best. Individual bands with higher Chern numbers are also less common and typically found in abstract models\cite{wang001,trescher000,yang000,zhang004,fang000}. It is therefore important to identify physically-relevant systems in which higher Chern numbers -- and QAH insulator phases with higher Chern numbers -- may be realized, with higher Chern numbers arising for individual bands being especially interesting. 

In light of earlier work showing that effective intersite SOC between $e_g$ orbitals can be induced perturbatively to yield a QAH state with Chern number $\pm1$ in double perovskite monolayers grown in the \{001\} crystallographic direction\cite{zhang002}, and earlier work on realizing QAH insulator phases with Chern number $\pm2$ in \{111\} double perovskite bilayers\cite{cook000}, we consider a tight-binding model for electrons in $t_{2g}$ orbitals describing double perovskite monolayers grown in the \{001\} crystallographic direction. We present evidence indicating that higher Chern numbers, including $\pm2$, $\pm4$, $\pm6$, and $\pm8$, can be realized for the lowest band of this model. This model is relevant to ferrimagnetic, half-metallic double perovskites, such as Ba$_2$FeReO$_6$, Sr$_2$FeMoO$_6$, Sr$_2$FeReO$_6$, Sr$_2$CrMoO$_6$, and Sr$_2$CrWO$_6$, where the large moments on Fe or Cr are treated classically. These results may be important given rapid developments in growing transition metal oxide (TMO) heterostructures\cite{ohtomo000,mannhart000,hwang000}. These results may furthermore be important to experimental realization of exotic fractional Chern insulator physics\cite{tang000, regnault000,neupert000} given correlations can vary greatly in strength in TMOs. 

\begin{figure}[t]
\includegraphics[width=8.5cm]{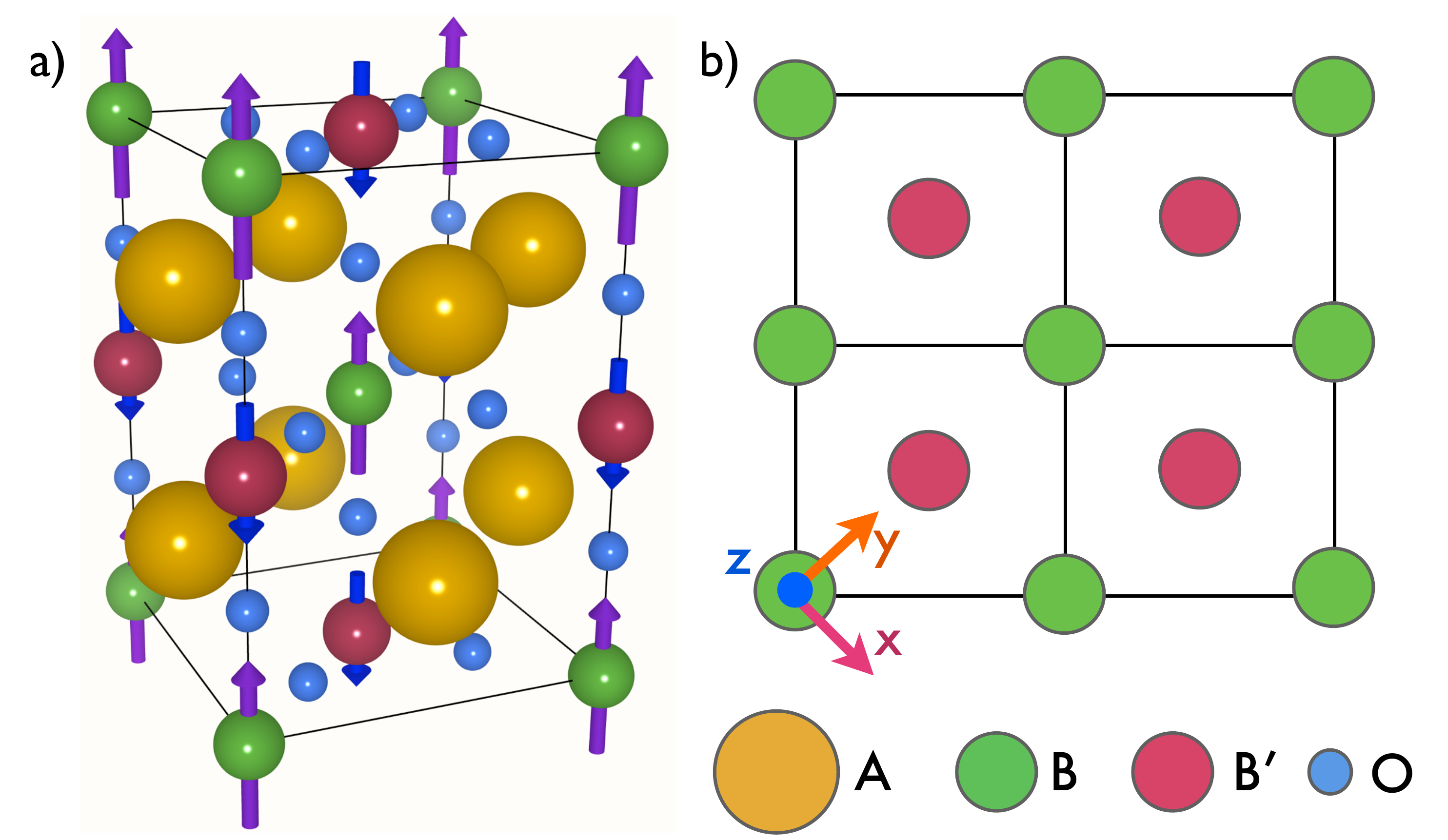}
\caption{\label{fig1} (a) Crystal structure of the bulk ordered double perovskite crystal
A$_2$BB'O$_6$ shown in perspective, with arrows depicting ferrimagnetic configuration of moments. (b) Double perovskite monolayer grown in the \{001\} crystallographic direction.  For this work, we consider half-metallic double perovskite Sr$_2$Cr WO$_6$ (SCWO) as an example.}
\end{figure}

The paper is organized as follows: In Section II, we show that a model describing a monolayer of SCWO grown in the \{001\} crystallographic direction, with the Cr moments ferromagnetically-aligned along high-symmetry directions, is topologically unstable, exhibiting point nodes or line nodes depending on the orientation of the ferromagnetic order.  We examine these numerical results using effective models describing the two lowest bands of the dispersion.  In Section III, we then demonstrate how different symmetry-breaking perturbations motivated by the effective models, including shifting of the $xz$ and $yz$ orbitals up out of the monolayer plane, tipping of the classical moments away from high-symmetry directions, and octahedral rotations can stabilize the band topology.  Stable topological phases realized through such symmetry breaking are characterized by Chern numbers for the lowest band of the model dispersion of $0$, $\pm1$, $\pm2$, $\pm4$, $\pm6$, and $\pm8$.  We further show it is possible to tune between some of these topological phases using weak, in-plane magnetic fields before going on to discuss future directions suggested by these results.

\section{Tight-Binding Model}

We use the double perovskite Sr$_2$CrWO$_6$ (SCWO) as an example in describing our model for half-metallic double perovskites, although the model also describes similar physics occurring in SFMO, SCMO, and BFRO. Strong Hund's coupling on Cr$^{3+}$ locks the 3d electrons into a local moment of $3\over 2$, large enough to be treated as a classical spin to good approximation. The 5d$^1$ electron on W$^{5+}$ hops onto Cr at a cost of charge-transfer energy $\Delta$. Pauli exclusion on Cr forces itinerant electrons from W to align antiparallel to the classical Cr moment, meaning itinerant electrons created at Cr sites are stripped of their spin degree of freedom. As a result of this physics, ferromagnetic order of Cr moments in bulk SCWO minimizes the ground state energy by favouring delocalization of the itinerant electrons in the system\cite{philipp000}. This Hamiltonian has been shown\cite{cook002} to capture the phenomenology of bulk Ba$_2$FeReO$_6$, quantitatively explaining its band dispersion\cite{jeon000}, saturation magnetization\cite{prellier000,teresa000}, spin and orbital polarizations\cite{azimonte000}, and spin dynamics observed using neutron scattering\cite{plumb000}. 

Here, we consider a monolayer of SCWO grown in the \{001\} crystallographic direction, which confines electrons to a square lattice (see Fig. 1). The W t$_2g$ orbitals transform as $L = 1$ angular momentum states, and experience local SOC, $- \lambda \vec{L} \cdot \vec{S}$, with $\lambda> 0$, leading to a low energy $J = 3/2$ quartet and a high energy $J = 1/2$ doublet. We can then write the model Hamiltonian as

\begin{align}
H &= \sum_{\langle i,j\rangle,\ell,\sigma}\left[ t^{ij}_{\ell} g_{\sigma}(j)d^{\dagger}_{i\ell \sigma} f^{}_{j \ell} + H.c. \right] + \Delta \sum_{i \ell} f^{\dagger}_{i \ell} f^{}_{i \ell} \\ \nonumber
&+\sum_{\langle \langle i,j\rangle \rangle,\ell,\sigma} \eta^{ij}_{\ell \ell'} d^{\dagger}_{i \ell \sigma} d^{}_{j \ell' \sigma}+ \sum_{\langle \langle i,j\rangle \rangle,\ell} \alpha^{ij} \eta^{ij}_{\ell \ell'} f^{\dagger}_{i \ell} f^{}_{j \ell' } \\ \nonumber &+i{\lambda \over 2}\sum_i \varepsilon_{\ell,m,n}\tau^n_{\sigma,\sigma'} d^{\dagger}_{i,\ell,\sigma}d_{i,m,\sigma'}+ H_{tet}
\end{align}

Operator $d$ ($f$) encodes electron annihilation on W (Cr), $i$ labels sites,
$\sigma$ labels spin, $\ell$ is the orbital
index differentiating between operators for electrons in each of the three $t_{2g}$ orbitals $xy$, $yz$, or $xz$, and $\varepsilon _{\ell m n}$ is the totally antisymmetric tensor. Defining classical Cr moment
orientation with the vector $\hat{F}= (\sin \theta \cos \phi, \sin \theta \sin \phi, \cos \theta)$, itinerant Cr electron spin orientation is characterized by projection constants $g_{\uparrow}(j) = \sin {\theta \over 2} e^{-i\phi_j /2}$ and $g_{\downarrow}(j) = -\cos {\theta \over 2} e^{i\phi_j /2}$. Matrix elements $t^{ij}$ correspond to intra-orbital Cr-W hopping amplitudes $t_{\pi}$ (in orbital plane) and $t_{\delta}$ (out of orbital plane).  $\eta^{ij}$ contains W-W intra-orbital hopping amplitudes $t'$ (in orbital plane) and $t''$ (out of orbital plane), as well as inter-orbital W-W hopping $t_m$.  $\alpha^{ij} \eta^{ij}$ contains Cr-Cr intra-orbital hopping amplitudes $\alpha t'$ (in orbital plane) and $\alpha t''$ (out of orbital plane), with inter-orbital Cr-Cr hopping neglected. We also include an additional symmetry-allowed tetragonal distortion in the direction normal to the monolayer plane, described by $H_{tet} = \sum_{i,\sigma} \left[ -\chi_{tet} \left( d^{\dagger}_{i,xz, \sigma}d^{}_{i,xz, \sigma} + d^{\dagger}_{i,yz, \sigma}d^{}_{i,yz, \sigma}  \right) \right]$.  Here, $\chi_{tet}<0$ corresponds to compressing the W oxygen octahedral cage. Finally, although not included explicitly in Eq. 1, $a$ is the length of nearest-neighbor W-Cr bonds and appears in later figures and equations.

For SCWO, our model captures the key energy scales:  Being a 3d/5d double perovskite, we expect energy scales similar to those of BFRO, save for $\Delta$. $\Delta$ can, however, be estimated from previous electron structure studies of SCWO to get $\Delta /t_{\pi} \sim 1.5$\cite{vaitheeswaran000}.  The full set of values being used for SCWO are $t_{\pi}=1$, $t_{\delta}=-0.1t_{\pi}$, $t'=-0.3t_{\pi}$, $t''=0.1t_{\pi}$, $\Delta=1.5 t_{\pi}$, $t_{m}=-0.1t_{\pi}$, $\lambda=2 t_{\pi}$, and $\alpha=0.5$. Since SCWO has one electron per unit cell, we neglect interactions.


We begin study of this model by computing a phase diagram as a function of W-W inter-orbital hopping $t_m$, tetragonal distortion $\chi_{tet}$, and Cr moment magnetic order in the ground state over reasonable regimes. Although the ground state magnetic order could be richer, ferromagnetic ordering of Cr moments is reasonably assumed in this paper as it maximizes delocalization of itinerant electrons in this model. We further restrict ourselves, initially, to orientations of these ferromagnetically-aligned Cr moments in distinct high-symmetry directions: For each point in the phase diagram, we compute the ground state energy for Cr moments aligned ferromagnetically in each of the five distinct high-symmetry directions (\{111\}, \{001\}, \{100\}, \{110\}, \{101\}).  For the lowest-energy orientation, we then compute the indirect gap between the lowest and second-lowest bands of the dispersion and the Chern number for the lowest band, $C(0)$. To compute Chern numbers, we primarily use a discrete lattice version of integration of the Berry connection desirable for its fast convergence\cite{souza000}.  We write the Chern number of the $n^{\rm{th}}$ band as
\begin{equation}
C(n) = {1\over 2\pi} \sum_{\bk} \rm{Im} \ln \left( A^n_{\bk, \hat{x}} A^n_{\bk + \hat{x}, \hat{y}} A^n_{\bk+\hat{x}+\hat{y},-\hat{x}} A^n_{\bk+\hat{y},-\hat{y}} \right)
\end{equation}
where the Berry connection is written in terms of eigenkets for the $n^{\rm{th}}$ band at point $\bk$ in the Brillouin zone, $| n \bk \rangle$, as $A^n_{\bk, \mu} = \langle n \bk | n \bk + \mu \rangle$.

\begin{figure}[t]
\includegraphics[width=9cm]{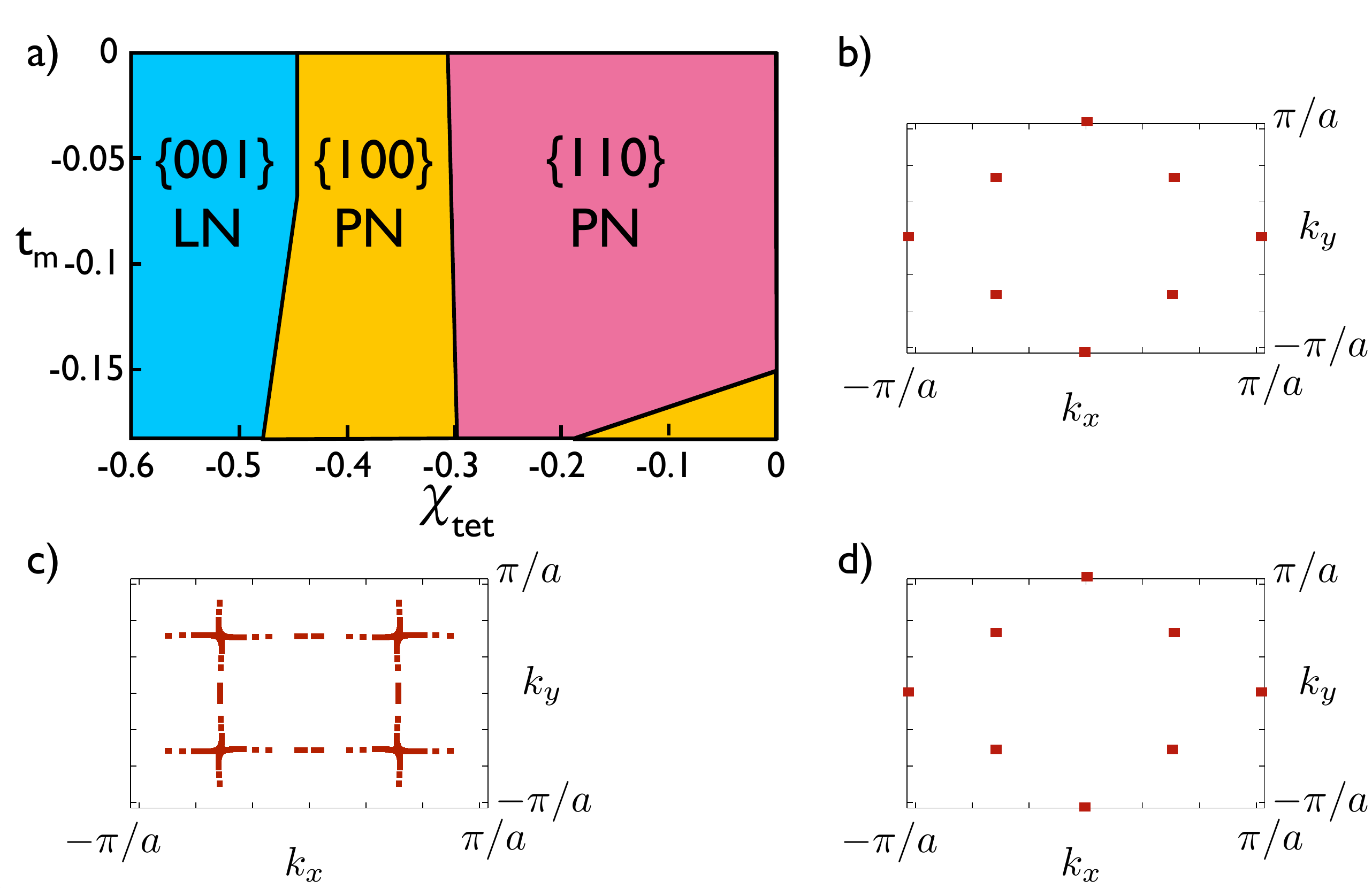}
\caption{\label{fig2} (a) Phase diagram for the full tight-binding model of the \{001\} monolayer with SCWO parameter set except for varying tetragonal distortion $\chi_{tet}$ (horizontal axis) and interorbital hopping $t_m$ (vertical axis). Ground state ferromagnetic orientations of Cr moments are given in \{\hspace{2mm}\}. Phases labeled PN are further characterized by point nodes due to intersections between the two lowest bands of the dispersion, and phases labeled LN exhibit line nodes due to intersections between the two lowest bands of the dispersion. Minimum negative indirect gap between the two lowest bands is greater than in magnitude than $0.7 t_{\pi}$ throughout the region shown, meaning the entire displayed region is metallic.(b) Band-touching plot inside \{100\} phase region showing point nodes between two lowest bands.  (c) Band-touching plot inside \{001\} phase region showing line nodes between two lowest bands.  (d) Band-touching plot inside \{110\} phase region showing point nodes between two lowest bands. Band-touching plots computed with step size $\Delta k = {0.01\over a}$ and maximum difference in energy between bands of $\Delta E = 0.001 t_{\pi}$.}
\end{figure}

This preliminary phase diagram is shown in Fig. \ref{fig2} (a).  For $\chi_{tet}=0$, we find Cr moments prefer to orient in the \{110\} crystallographic direction.  As $\chi_{tet}$ becomes more negative, the \{100\} phase dominates, and then the \{001\} phase is eventually preferred.  Chern number $C(0)$ is not shown for these phases, as each is actually topologically unstable as a result of band-touchings.  Instead, band-touchings are characterized for each phase:  Fig. \ref{fig2} (b), (c), and (d) depict numerical calculation of band-touchings, generated by plotting points in momentum-space at which the lowest and second lowest bands of the dispersion differ in energy by less than 0.001$t_{\pi}$.  Fig. \ref{fig2} (b) and (d) thus show point nodes present for the \{100\} and \{110\} phases, while Fig. \ref{fig2} (c) shows the line nodes present in the \{001\} phase.  The system is metallic everywhere in the phase diagram, with an indirect gap greater in magnitude than $0.7t_{\pi}$ throughout the region of phase space shown.

\begin{figure}[t]
\includegraphics[width=8.5cm]{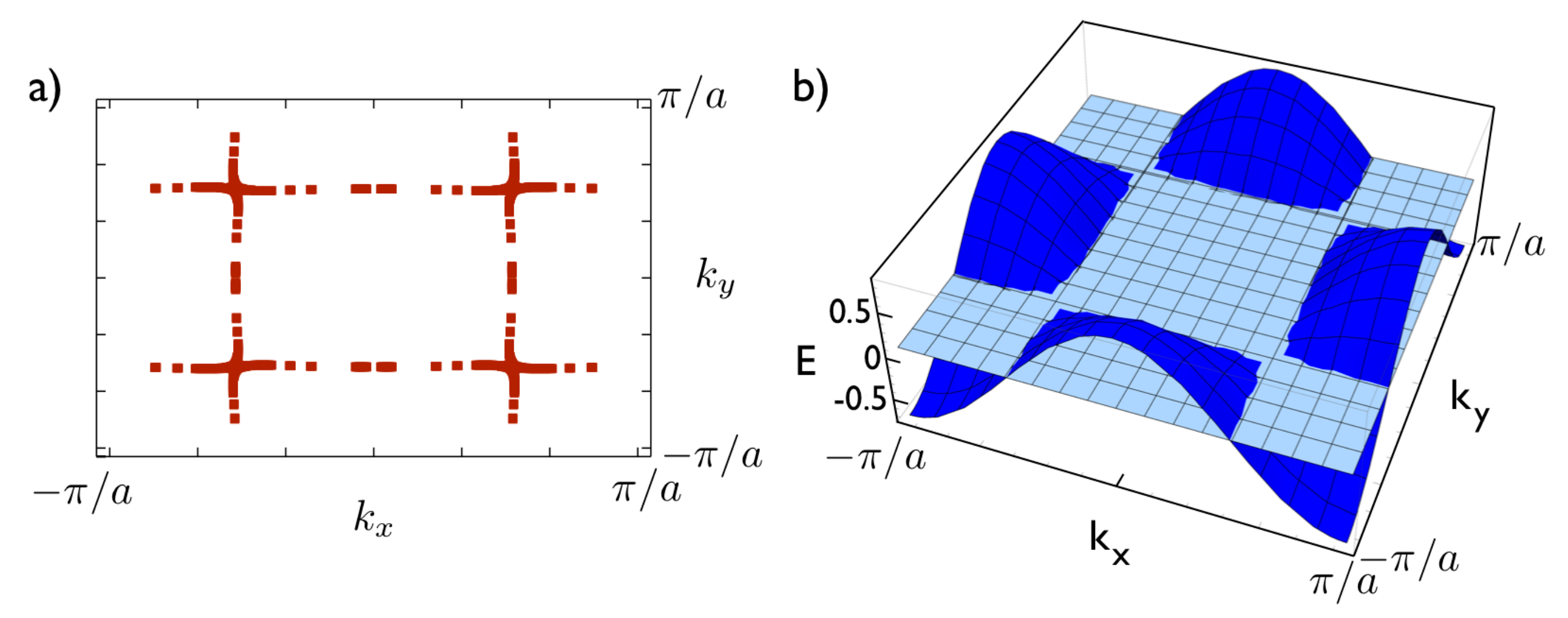}
\caption{\label{fig3} (a) Plot showing band-touchings in \{001\} phase between two lowest bands of the full 9-band tight-binding model for $t_m=-0.1t_{\pi}$, $\chi_{\rm{tet}}=0$.  (b) Band structure of $H^{\{001\}}_{\rm{eff}}$ for $\chi_{\rm{tet}}=-0.18$, $t'=0.3$, and $z=0.1$, showing similar line node structure.}
\end{figure}
We gain some understanding of the origin of the point and line node degeneracies observed in the full $9$-band model from effective models written by projecting the full Hamiltonian onto the two lowest-energy eigenstates to create effective $2$-band models as done in earlier work\cite{cook000}.  For weak tetragonal distortion, we know the forms of these two lowest states of the full model: They are two from the $J = {3 \over 2}$ multiplet, $|J=-{3\over 2} \rangle$ and $|J=-{1\over 2} \rangle$.  For Cr moments aligned ferromagnetically in the \{001\} direction, these are

\begin{eqnarray}
|J_z=-3/2\rangle &=& |L=-1\rangle |S=-1/2\rangle = {1\over \sqrt{2}}\left( |xz\rangle - i |yz \rangle \right) |\downarrow \rangle \nonumber \\
|J_z=-1/2\rangle &=& \left(S^+ + L^+\right) |L=-1\rangle |S=-1/2\rangle  \nonumber \\
&=& {1\over \sqrt{6}}\left( |xz\rangle - i |yz \rangle \right) |\uparrow \rangle + \sqrt{{2\over 3}} |xy\rangle |\downarrow \rangle \nonumber
\end{eqnarray}

Writing $\cos(k_x a)$ as $c_x$, $\cos(k_y a)$ as $c_y$, $\sin(k_x a)$ as $s_x$, and $\sin(k_y a)$ as $s_y$, we have
\begin{equation}
H^{\{001\}}_{\rm{eff}}(k) = \left( \begin{matrix} -{8\over 3} t' c_x c_y - {\chi_{\rm{tet}} \over 3} & 0 \\  0 &-\chi_{\rm{tet}} - z \end{matrix} \right)
\end{equation}

where $z$ is the difference in energy between the two states of the $J={3\over 2}$ manifold due to Zeeman splitting. We see this Hamiltonian, possessing non-zero terms only on the diagonal, exhibits line nodes. We further find good qualitative agreement between lines nodes of the full model shown in Fig. \ref{fig3} (a) and those of $\{001\}$ effective model shown in Fig. \ref{fig3} (b) for zero tetragonal distortion, where the full model's low-energy states are reasonably-assumed to be two from the $J=3/2$ manifold.

With Cr moments instead aligned ferromagnetically in the \{100\} direction, the $|J=-{3\over 2} \rangle$ and $|J=-{1\over 2} \rangle$ states instead take the form
\begin{eqnarray}
|J_x=-3/2\rangle &=& |L=-1\rangle |S=-1/2\rangle = {1\over \sqrt{2}}\left( |xz\rangle - i |xy \rangle \right) |\downarrow \rangle \nonumber \\
|J_x=-1/2\rangle &=& \left(S^+ + L^+\right) |L=-1\rangle |S=-1/2\rangle  \nonumber \\
&=& {1\over \sqrt{6}}\left( |xz\rangle - i |xy \rangle \right) |\uparrow \rangle + \sqrt{{2\over 3}} |yz\rangle |\downarrow \rangle \nonumber
\end{eqnarray}
Projection of hoppings in the full model onto these two states yields the following effective Hamiltonian:
\begin{equation}
H^{\{100\}}_{\rm{eff}}(k) = \left( \begin{matrix} -{8\over 3} t' c_x c_y - {5\over 6} \chi_{\rm{tet}}& {-4\over \sqrt{3}}t_m s_x s_y\\ {-4\over \sqrt{3}} t_m s_x s_y &-2t'c_x c_y -{1\over 2} \chi_{\rm{tet}} - z \end{matrix} \right)
\end{equation}
We see this effective model instead exhibits point nodes, in agreement with the full model.

Finally, with Cr moments aligned ferromagnetically in the \{110\} direction, the $|J=-{3\over 2} \rangle$ and $|J=-{1\over 2} \rangle$ states can be written as
\begin{eqnarray}
|J_{110}=-3/2\rangle &=& |L=-1\rangle |S=-1/2\rangle \nonumber \\
&=& \left( {-i\over \sqrt{2}}|xy\rangle - {1\over 2} |yz\rangle+ {1\over 2} |xz \rangle \right) |\downarrow \rangle \\
|J_{110}=-1/2\rangle &=& \left(S^+ + L^+\right) |L=-1\rangle |S=-1/2\rangle  \nonumber \\
&=& {1\over \sqrt{3}}\left( {-i\over \sqrt{2}}|xy\rangle - {1\over 2} |yz\rangle+ {1\over 2} |xz \rangle \right) |\uparrow \rangle \nonumber \\
&+& {1\over \sqrt{3}} \left(  |yz\rangle + |xz\rangle \right)|\downarrow \rangle
\end{eqnarray}

\begin{equation}
H^{\{110\}}_{\rm{eff}}(k) = \left( \begin{matrix} A & 0 \\ 0 & B \end{matrix} \right)
\end{equation}
where 
\begin{eqnarray}
A &=& -{2\over 3} t' c_x c_y - 2t_m s_x s_y - {5\chi_{\rm{tet}}\over 6} \\
B &=& -2t' c_x c_y -2t_m s_x s_y-{\chi_{\rm{tet}}\over 2} - z 
\end{eqnarray}

In the $\{110\}$ phase, the effective model fails to capture the point nodes observed in the full model. It is plausible that W-Cr hopping, higher bands, or other physics of the full model neglected in the effective model might be responsible for the additional symmetry-breaking.

\section{Stabilizing topological phases}

\subsection{Shifting $xz$ and $yz$ orbitals up out of monolayer plane}

\begin{figure}[t]
\includegraphics[width=8cm]{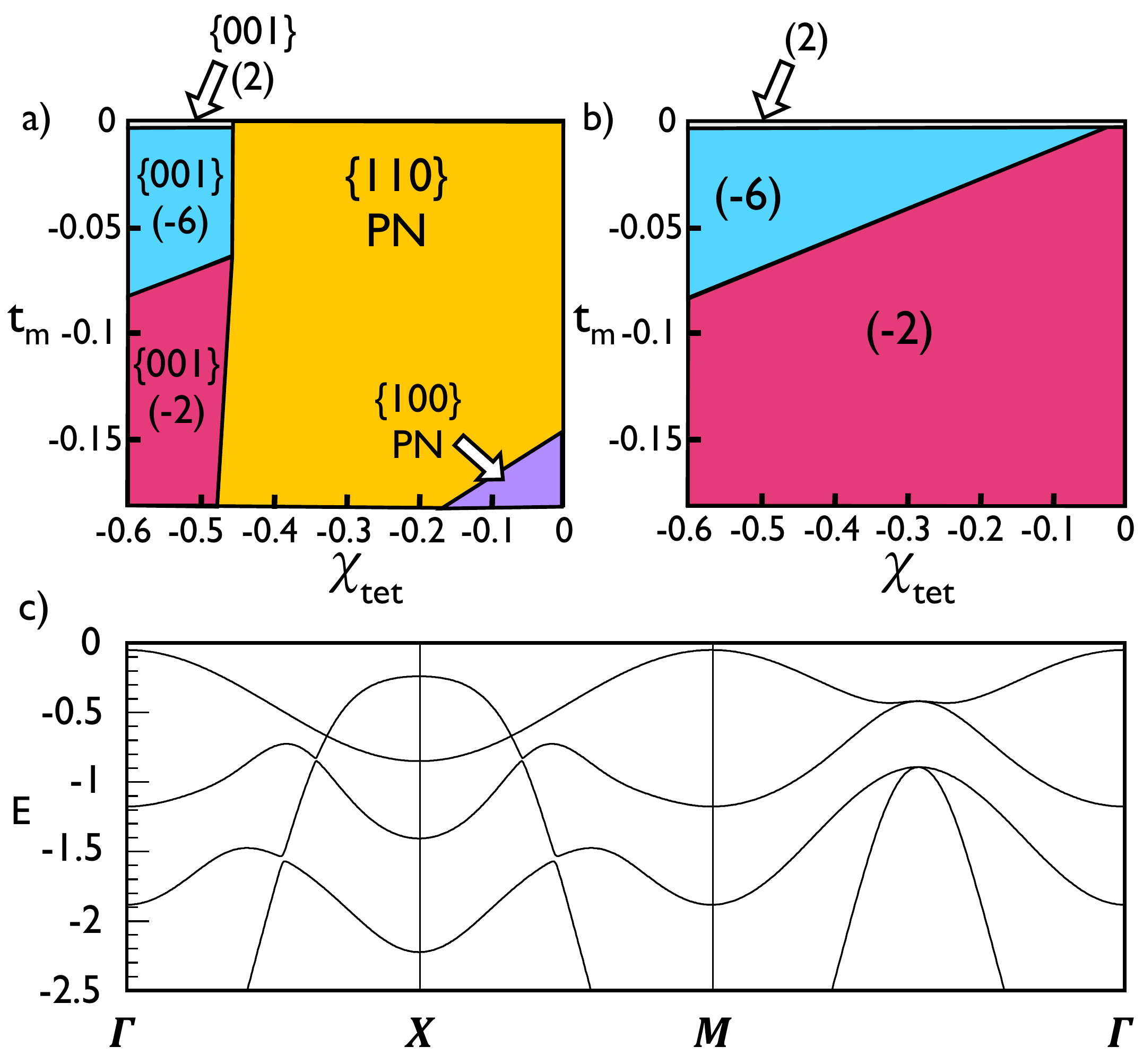}
\caption{\label{fig7} Phase diagrams for the full tight-binding model with SCWO parameter set -- but with varying tetragonal distortion $\chi_{tet}$ (horizontal axis) and interorbital hopping $t_m$ (vertical axis) -- as well as xz/yz orbital shift included ($t_{mp}=0.01 t_{\pi}$) showing (a) the Chern number of the lowest band, $C(0)$, for various global orientations of Cr moments in the ground state. Ground state ferromagnetic orientations of Cr moments are given in \{\}. $C(0)$ for each phase is given in round brackets, (). PN denotes a phase in which point nodes exist due to intersection of the two lowest bands. (b) Phase diagram with with Cr moments ferromagnetically-aligned but fixed to point in the \{001\} direction rather than the orientation in the ground state. All phases shown here in (a) and (b) are metallic, with a negative indirect gap between the two lowest bands of the model of at least $\sim 0.7 t_{\pi}$ everywhere shown. (c) Dispersion of the full tight binding model plotted in units of $t_{\pi}$ along high-symmetry lines in the Brillouin zone near the phase boundary between the $C(0) = -6$ region and $C(0) = -2$ region (at $t_m = -0.075 t_{\pi}$ and $\chi_{\rm{tet} } = -0.55 t_{\pi}$) in (a) and (b). Only the four bands of the dispersion lowest in energy are visible, corresponding to a positive indirect gap between the $4^{\rm{th}}$ and $5^{\rm{th}}$ lowest bands in energy. The topological phase transition corresponds to closing of the band gap at four momenta $\left( k_x, k_y\right) = \left( { \pm \pi \over 2a}, {\pm \pi \over 2a} \right)$ and $\left( k_x, k_y\right) = \left( { \pm \pi \over 2a}, {\mp \pi \over 2a} \right)$, with each band-touching transferring Chern number of $1$ corresponding to a total change in the Chern number of the lowest band of $4$ at the phase transition.}
\end{figure}

From the derivation of the effective two-band model for the Hamiltonian $H^{\{001\}}$, we see the lack of inter-orbital W-W hopping between $xz/yz$ and $xy$ orbitals results in line node degeneracies.  This motivates adding a new term to the full model, corresponding to a small shift of the $xz$ and $yz$ orbitals in the $+\hat{z}$ direction, up out of the monolayer plane, to allow for a small amount of inter-orbital $xz \leftrightarrow xy$ and $yz \leftrightarrow xy$ hopping.  Such a shift might be induced by growing the monolayer on a substrate, or in heterostructure.  This new term, $H^{\rm{shift}}_{\sigma}(k)$, can be written in momentum space as
\begin{align}
H^{\rm{shift}}_{\sigma}(k)=2i t_{mp} d_{k,xy_{\sigma}}^{\dagger}d_{k,yz_{\sigma}}\left[ \sin(k_x a + k_y a) \right. \\ \nonumber
\left. + \sin(k_x a - k_y a) \right] + h.c. \\ \nonumber
+ 2i t_{mp} d_{k,xy_{\sigma}}^{\dagger}d_{k,xz_{\sigma}}\left[ \sin(k_x a + k_y a) \right. \\ \nonumber
\left. - \sin(k_x a - k_y a) \right] + h.c.
\end{align}
and here, $t_{mp}$ is chosen to be a small value of $0.01 t_{\pi}$.  Again, we compute a phase diagram by determining which one of the five distinct, high-symmetry orientations the Cr moments prefer to align along ferromagnetically in the ground state, and then computing the Chern number $C(0)$ of the lowest band for that ground state magnetic order.  This phase diagram is shown in Fig. \ref{fig7}(a).


The phase diagram shows topologically unstable \{110\} and \{100\} phases.  The topological instability of these phases can be understood from the effective \{110\} and \{100\} models, which show the newly-introduced $xz \leftrightarrow xy$ and $yz \leftrightarrow xy$ hoppings are not expected to stabilize these phases.  The shift, however, does indeed stabilize the \{001\} phase as expected, yielding three regions:  There is a $C(0)=2$ for $t_m=0$, a $C(0)=-2$ region for large $t_m$, and also a $C(0)=-6$ region.  The system is again everywhere metallic, with an indirect gap greater than $0.7 t_{\pi}$ in magnitude at the least. Computing the dispersion near the boundary between the $C(0)=-2$ and $C(0)=-6$ regions, we find that Chern number $4$ is being transferred between the two lowest bands by four Dirac-like band-touchings occurring at $\left( k_x, k_y\right) = \left ({\pi \over 2a}, {\pi \over 2a} \right)$, $\left (-{\pi \over 2a}, -{\pi \over 2a} \right)$, $\left ({\pi \over 2a}, -{\pi \over 2a} \right)$, and $\left (-{\pi \over 2a}, {\pi \over 2a} \right)$, as shown in Fig. \ref{fig7}(c).

Forcing the Cr moments to align in the \{001\} crystallographic direction for all $t_m$ and $\chi_{tet}$, we generate the related phase diagram shown in Fig. \ref{fig7}(b).  We see that the $C(0)=-6$ region extends across the phase diagram to almost $\chi_{tet}=0$ in a large triangle.  This can be understood using the effective two-band model for the \{001\} phase.  With the new shift term included, $H_{\{001\}}$ becomes

\begin{equation}
H^{\{001\}}_{\rm{eff}}(k) = \left( \begin{matrix} -{8\over 3} t' c_x c_y - {\chi_{\rm{tet}} \over 3} & {4\over \sqrt{3}} t_{mp} \left[s_x c_y + i c_x s_y \right]\\ {4\over \sqrt{3}} t_{mp} \left[ s_x c_y - i c_x s_y \right] &-\chi_{\rm{tet}} - z \end{matrix} \right)
\end{equation}

We can compute the Chern numbers of this two-band model.  At zero tetragonal distortion, $C(0) = 2$.  As a function of decreasing $\chi_{\rm{tet}}$ starting from zero tetragonal distortion, $C(0)$ jumps from $2$ to $-2$ via four Dirac-like band touchings.  Thus, although the effective model does not capture the absolute Chern numbers observed in the full model phase diagram, it does possess a relative drop in Chern number by 4 with increasing tetragonal compression.  This topological phase transition occurs in the effective model on a line defined by Zeeman splitting $z$ proportional to tetragonal distortion $\chi_{tet}$. This is similar to what is found in the corresponding phase diagram for the full model, where a topological phase boundary occurs on a line $t_m \propto \chi_{tet}$. We can understand this similarity between the effective model and full model to a certain extent as follows: In the monolayer model, there exists only $xz$/$yz$ interorbital hopping for finite $t_m$ and $t_{mp}=0$, meaning that a key effect of $t_m$ is changing the energy of itinerant electrons in $xz$ and $yz$ orbitals via delocalization. For \{001\} orientation of the ferromagnetically-aligned Cr moments, Zeeman splitting also effectively changes the energy of electrons in $xz$/$yz$ orbitals relative to those in $xy$, by changing the energy of the $|J_z = -3 / 2 \rangle$ state relative to the $|J_z = - 1 / 2 \rangle$ state. The agreement between the effective model and full model as to the form of this phase boundary further supports our understanding of the emergence of topologically-nontrivial phases in the full model.

\begin{figure}[t]
\includegraphics[width=8.5cm]{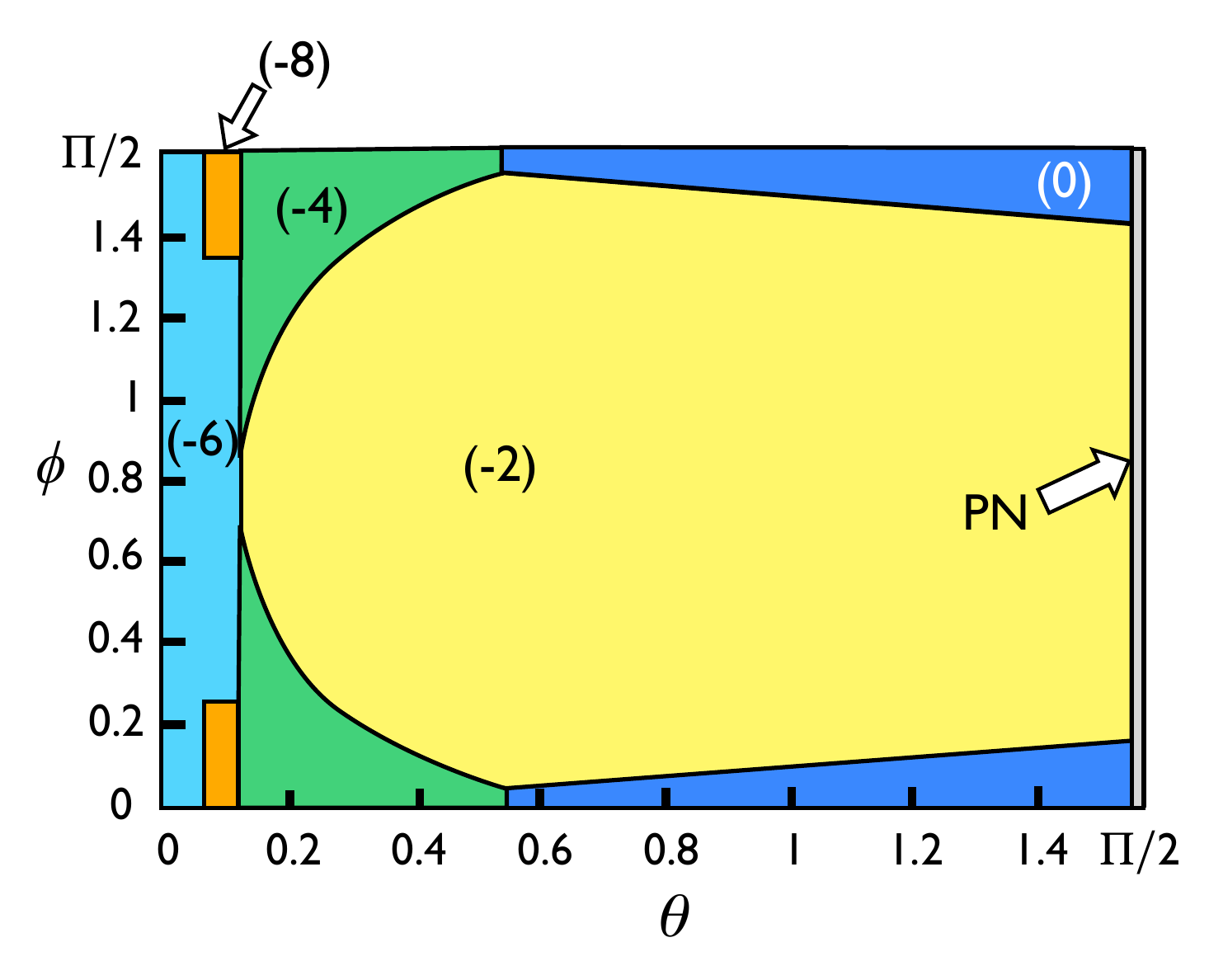}
\caption{\label{fig8_2} Phase diagram of the full tight-binding model with SCWO parameter set (but with $t_m=-0.03 t_{\pi}$, $\chi_{\rm{tet}}=-0.55 t_{\pi}$) when $xz$ and $yz$ orbitals are shifted up out of the monolayer plane ($t_{mp}=0.01 t_{\pi}$). Chern number of the lowest band $C(0)$ is computed as a function of global Cr moment orientation, given by angles $\theta$ and $\phi$, with $C(0)$ for each phase shown in round brackets. PN denotes a phase in which point nodes exist due to intersection of the two lowest bands. Phase diagram is everywhere metallic, with a negative indirect gap between the two lowest bands of the model on the order of $t_{\pi}$.}
\end{figure}

Choosing a point in the full model phase diagram where Cr moment orientation in the ground state is $\{001\}$, we compute Chern number $C(0)$ as a function of Cr moment orientation angles $\theta$ and $\phi$, corresponding to tilting the Cr moments with an applied in-plane magnetic field.  The resulting phase diagram is shown in Fig. \ref{fig8_2}.  This phase diagram is rich and highly-structured, displaying Chern numbers for the lowest band of $0$, $-2$, $-4$, $-6$, and $-8$ in a pattern symmetric about $\phi= {\pi \over 4}$.  Thus, it is also possible to place a system described by this model in the \{001\} phase via sufficient tetragonal compression of the oxygen octahedral cages and then apply a small in-plane magnetic field on the order of $0.1 t_{\pi}/ \mu_B$, to transition the system from the $C(0)=-6$ phase into other topological phases, which may be important for further investigation of these phases.

\subsection{Tipping classical moments away from high-symmetry directions}

\begin{figure}[t]
\includegraphics[width=8.5cm]{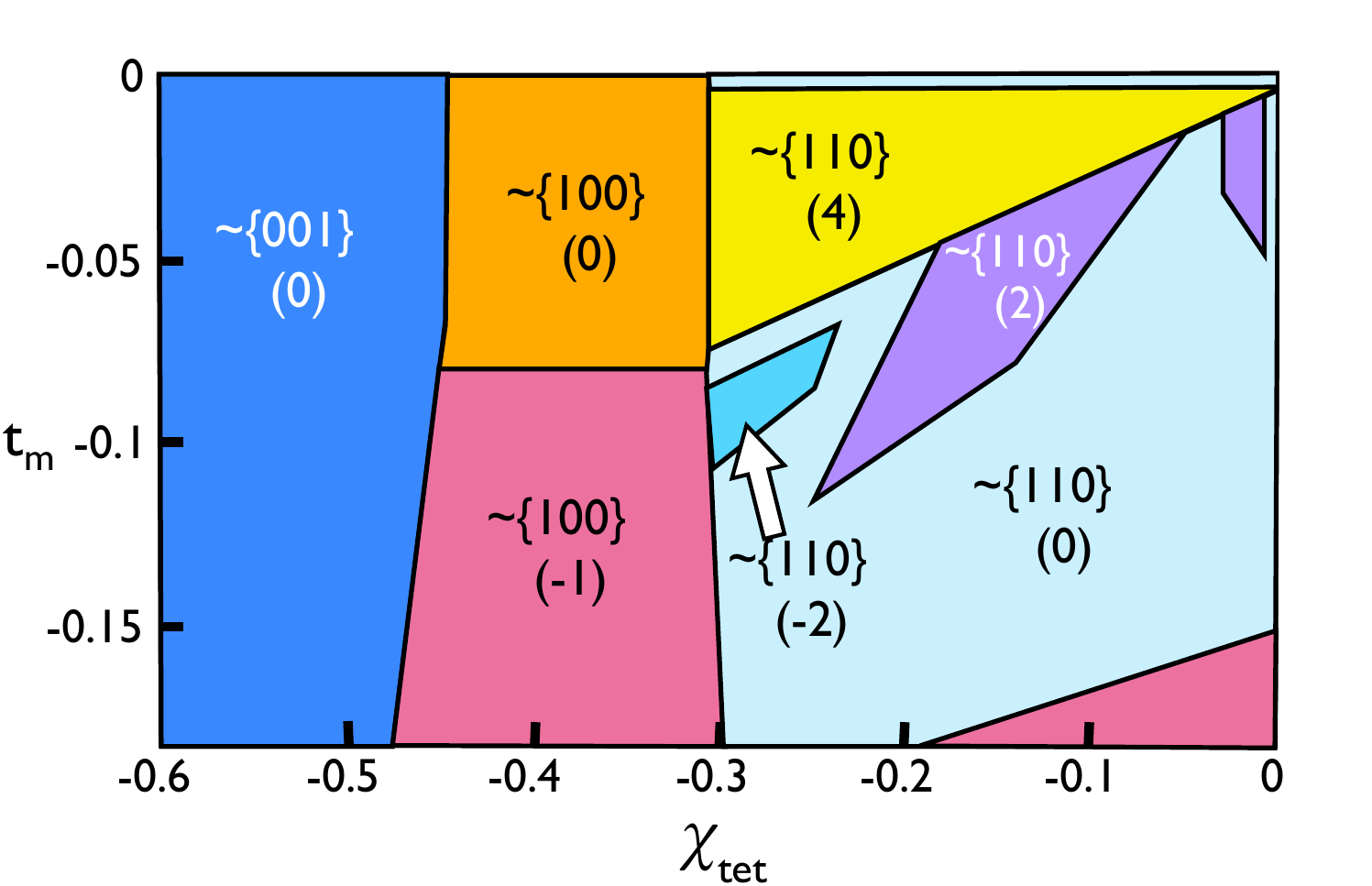}
\caption{\label{fig4} Phase diagram for the full tight-binding model with SCWO parameter set -- but with varying tetragonal distortion $\chi_{tet}$ (horizontal axis) and interorbital hopping $t_m$ (vertical axis) -- with small tilting of Cr moments away from high-symmetry directions.  Here, $\sim$\{001\} denotes the phase with ground state Cr moment orientation $\theta=0.01$ and $\phi=0.01$, $\sim$\{100\} denotes phase with $\theta=\pi/2+0.01$ and $\phi=0.01$, and $\sim$\{110\} denotes the phase with $\theta=\pi/2+0.01$ and $\phi=\pi/4+0.01$. The value of the Chern number for the lowest band, $C(0)$, in each phase is given in round brackets.  Phase boundaries are approximate. All phases shown here are metallic, with a negative indirect gap between the two lowest bands of the model of at least $\sim 0.7 t_{\pi}$ everywhere shown.}
\end{figure}

The effective models also suggest we can stabilize the band topology by tipping the classical Cr moments away from high-symmetry directions by small amounts rather than adding an $xz$/$yz$ orbital shift.  Taking $d \phi =0.01$ radians and $d \theta = 0.01$ radians, we regenerate the phase diagram shown in Fig. \ref{fig2}, but now compare ground state energies for Cr moments ferromagnetically-aligned along $\sim \{111\}=\{111\}+d \phi + d \theta$, $\sim \{001\}=\{001\}+d \phi + d \theta$, $\sim \{100\} = \{100\}+d \phi + d \theta$, $ \sim \{110\} = \{110\}+d \phi + d \theta$, and $ \sim \{101\} = \{101\}+d \phi + d \theta$ instead of \{111\}, \{001\}, \{100\}, \{110\}, \{101\}.  This phase diagram is shown in Fig. \ref{fig4}.

\begin{figure}[t]
\includegraphics[width=8.5cm]{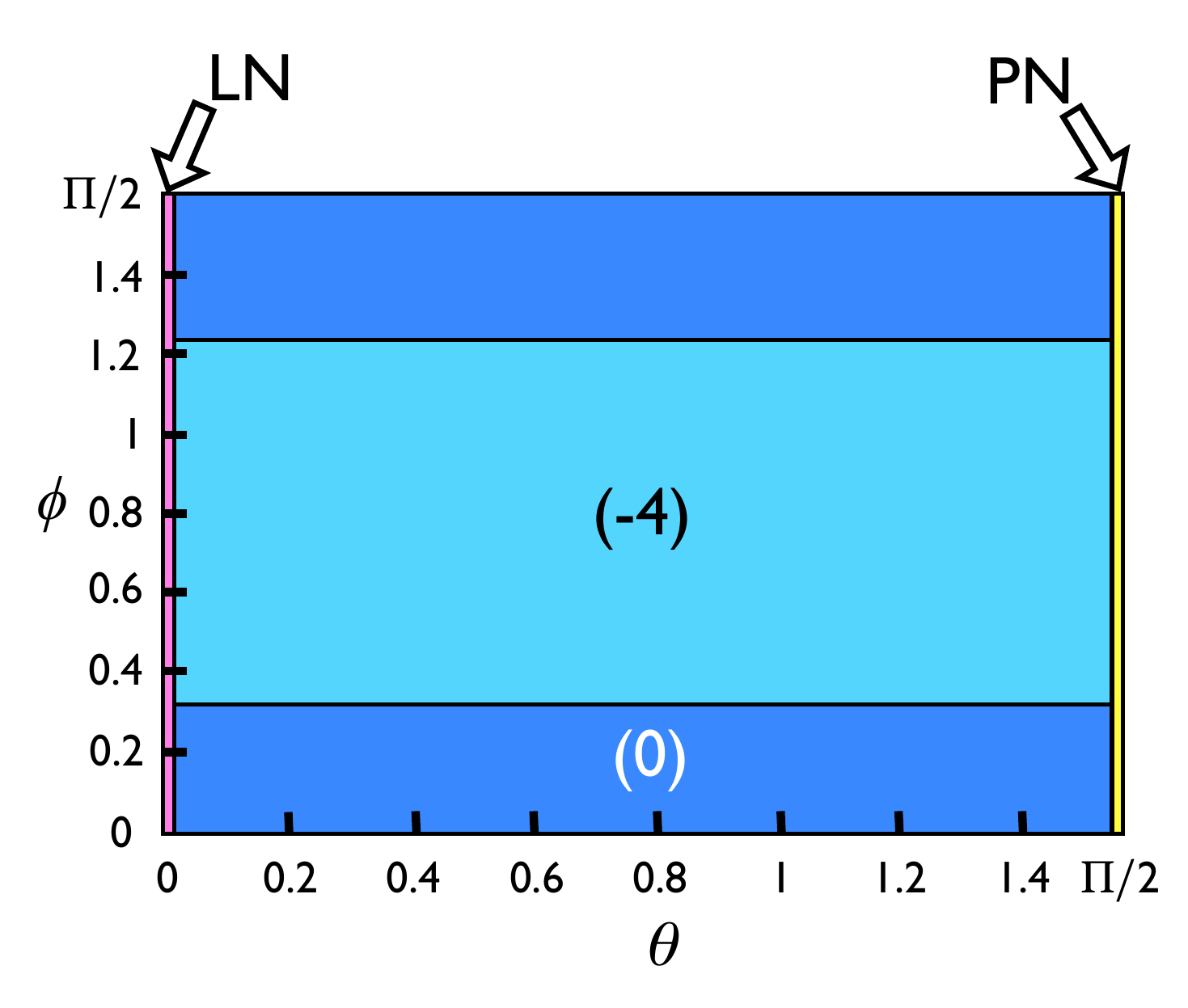}
\caption{\label{fig5} Phase diagram of the full tight-binding model with SCWO parameter set but with the following changes: $t_m=-0.03 t_{\pi}$, $\chi_{\rm{tet}}=-0.55 t_{\pi}$. (There is also no xz/yz orbital shift, so $t_{mp}=0$.) Chern number of the lowest band, $C(0)$, is computed as a function of ferromagnetically-aligned Cr moment orientation given by $\theta$ and $\phi$, with $C(0)$ shown in round brackets. Where $C(0)$ is unstable, LN denotes a phase in which line nodes exist due to intersection of the two lowest bands and PN denotes a phase in which point nodes exist due to intersection of the two lowest bands. Phase diagram is everywhere metallic, with a negative indirect gap between the two lowest bands of the model on the order of $t_{\pi}$.}
\end{figure}

Such calculation yields a phase diagram displaying a variety of Chern numbers for the lowest band of the dispersion, including $-1$, $-2$, $+2$, and $+4$.  Observing that the $C(0)=+4$ region extends to much more negative $\chi_{tet}$ if the system is forced to remain in the $\sim \{110\}$ phase rather than the $\sim \{001\}$ phase, we compute Chern number $C(0)$ as a function of Cr moment orientation angles $\theta$ and $\phi$ in the region where the $\sim \{001\}$ phase is preferred, but $C(0)=+4$ is expected if $\sim \{110\}$ is forced.  The system remains metallic everywhere, with the indirect gap being greater than $0.7 t_{\pi}$ at the least.  This phase diagram is shown in Fig. \ref{fig5}.

For $\theta=0$, the two lowest bands of the dispersion generate line nodes, while, at $\theta = {\pi \over 2}$, the two lowest bands of the dispersion generate point nodes.  For intermediate $\theta$, however, a region with $C(0)=-4$ exists in a band about $\phi= {\pi \over 4}$, bounded above and below by $C(0)=0$ regions, with Chern number of $4$ being transferred between the two lowest bands of the dispersion at this phase transition by four Dirac-like band-touchings occurring at $\left( k_x, k_y\right) = \left ({\pi \over 2a}, {\pi \over 2a} \right)$ and three other symmetry-related points in the Brillouin zone $\left (-{\pi \over 2a}, -{\pi \over 2a} \right)$, $\left ({\pi \over 2a}, -{\pi \over 2a} \right)$, and $\left (-{\pi \over 2a}, {\pi \over 2a} \right)$. These results indicate higher Chern numbers for the lowest band of the dispersion are possible even without the $xz$/$yz$ orbital shift. They furthermore suggest it is possible to push the system into a \{001\} ground state with sufficient tetragonal compression, and then tip the moments into either the $C(0)=-4$ or $C(0)=0$ regions by applying in-plane magnetic fields, which may again be useful for future investigation of these topologically-nontrivial phases. 

\subsection{Octahedral rotations}
We also consider two kinds of staggered octahedral rotations that may be relevant to experimental realizations of double perovskite monolayers and explore the extent to which they can stabilize exotic topological phases. The hopping parameters are obtained using the Slater-Koster\cite{slater000} method. To fully determine nearest-neighbour hoppings, however, we require a value for $t_{\sigma}$.  It is standard to use a ratio of $t_{dd\sigma}$:$t_{dd\pi}$ of $-{3\over 2}$: $1$\cite{andersen000,finnis000}, so we set $t_{\sigma} = -{3\over 2} t_{\pi}$.

\begin{figure}[t]
\includegraphics[width=8.5cm]{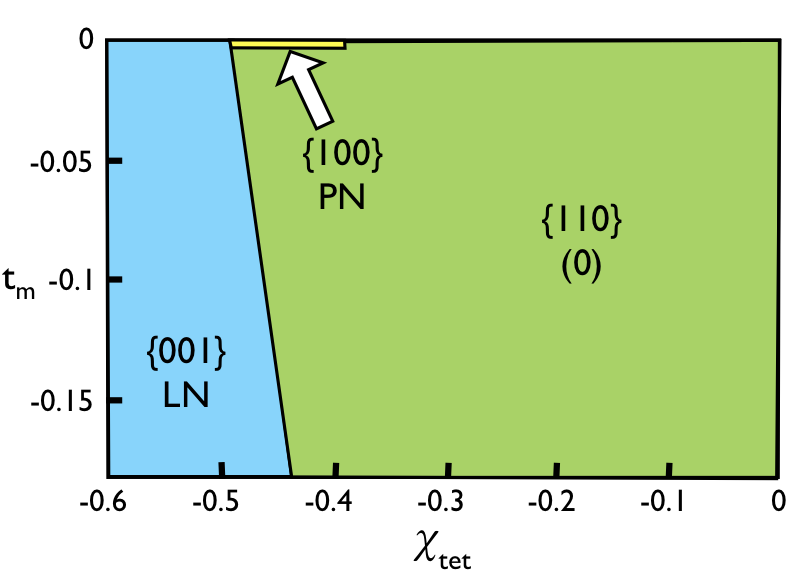}
\caption{\label{fig8} Phase diagram for the full tight-binding model with SCWO parameter set -- but with varying tetragonal distortion $\chi_{tet}$ (horizontal axis) and interorbital hopping $t_m$ (vertical axis) -- showing Chern number for the lowest band, $C(0)$ in round brackets for ground state orientations of ferromagnetically-aligned Cr moments.  Here, all phases are metallic, with a negative indirect gap between the two lowest bands of at least $\sim 0.7t_{\pi}$ everywhere shown. Ground state orientations of ferromagnetically-aligned Cr moments are given in \{\}. The \{110\} phase is topologically stable, with Chern number for the lowest band $C(0)=0$, while the \{001\} phase is topologically unstable. LN denotes a phase in which line nodes exist due to intersection of the two lowest bands. PN denotes a phase in which point nodes exist due to intersection of the two lowest bands.}
\end{figure}

First, we re-compute the phase diagram shown in Fig. \ref{fig2}, but for the case of small staggered octahedral rotations about the $\hat{z}$-axis ($\hat{c}$-axis) of 5 degrees.  This phase diagram is shown in Fig. \ref{fig8}.  The \{110\} appears to have spread, largely eliminating the \{100\} phase, but the \{001\} region remains roughly as large as before, though slightly different in shape.  As expected from the effective models, the imaginary $xz \leftrightarrow yz$ inter-orbital hopping introduced by such rotations stabilizes the \{110\} phase, but does not eliminate the line nodes previously observed in the \{001\} phase nor the point nodes in the small \{100\} phase region existing for $t_m=0$. 

It is natural to ask whether these octahedral rotations might affect the phase diagram shown in Fig. \ref{fig5}. Given the wide range in $\theta$ and $\phi$ over which the Chern number $4$ phase occurs in this phase diagram, however, it is likely that this phase would remain present and its boundaries largely unaffected by small octahedral rotations of this type. Thus, although octahedral rotations about the $\hat{c}$-axis do not appear to yield further exotic Chern numbers for the lowest band, they are not expected to destroy the exotic topology observed in the model without an additional shift of the $xz$/$yz$ orbitals up out of the monolayer plane.

\begin{figure}[t]
\includegraphics[width=8.5cm]{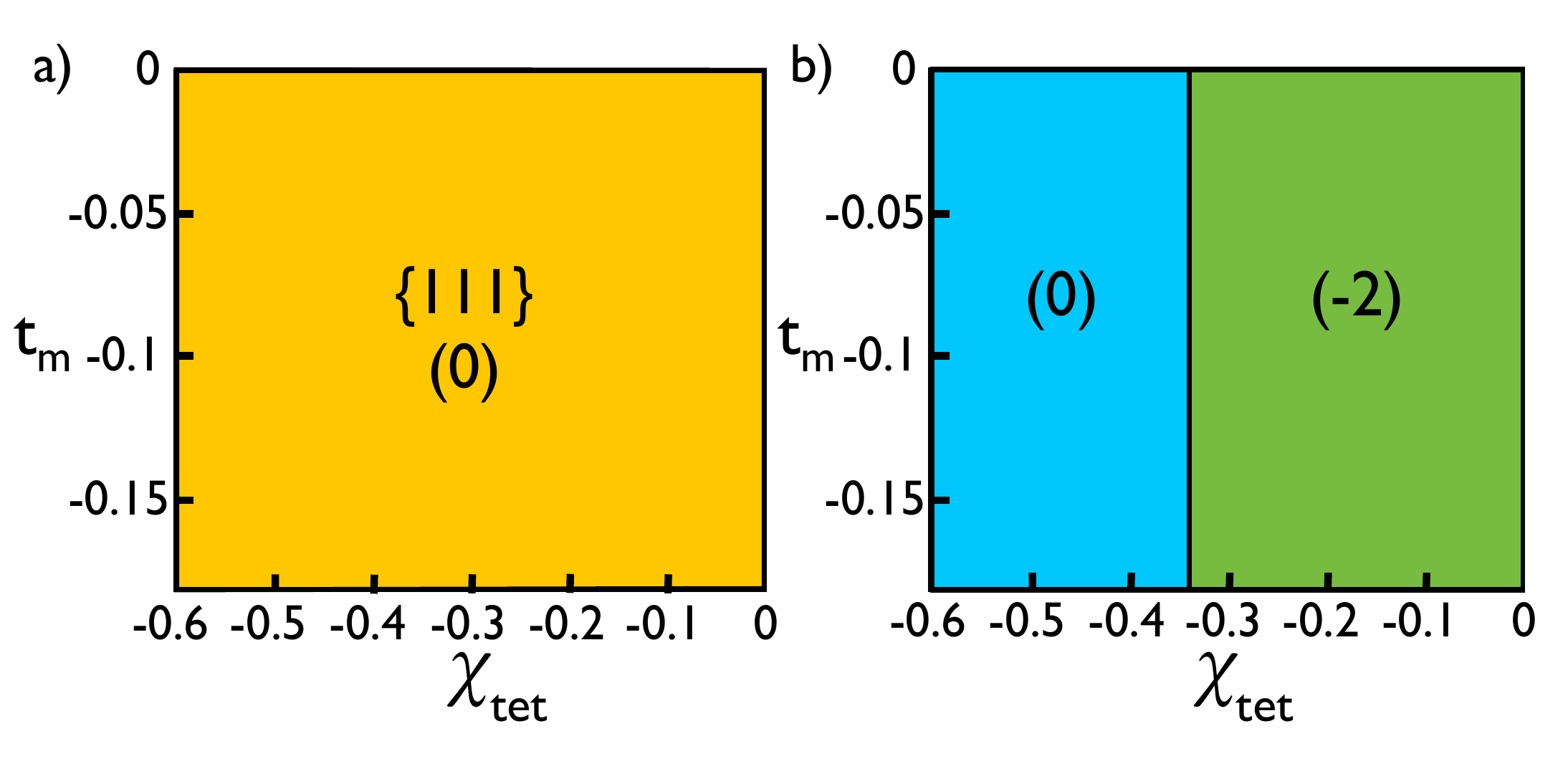}
\caption{\label{fig10} Phase diagrams for the full tight-binding model with SCWO parameter set, but with varying tetragonal distortion $\chi_{tet}$ (horizontal axis) and interorbital hopping $t_m$ (vertical axis). Subfigure (a) shows the Chern number in round brackets for the lowest band of the full model, $C(0)$, when Cr moments are ferromagnetically-aligned in the ground state orientation, \{111\}. Subfigure (b) shows the Chern number for the lowest band of the full model, $C(0)$, in round brackets for Cr moments aligned in the \{001\} direction. All phases shown here in (a) and (b) are metallic, with a negative indirect gap between the two lowest bands of the model of at least $\sim 0.7 t_{\pi}$ everywhere shown.}
\end{figure}

We next recompute Fig. \ref{fig2}(a) for the case of staggered octahedral rotations about the $\hat{x}+\hat{y}$ axis, or $\hat{b}$ axis, of roughly $5$ degrees.  The relevant phase diagram is shown in Fig. \ref{fig10}(a). Of the five distinct, high-symmetry ferromagnetic orderings of Cr moments considered previously, the lowest energy phase with such rotations now has \{111\}-aligned ferromagnetic order, exhibiting Chern number 0 for the lowest band.  If we force the Cr moments to align along the \{001\} crystallographic direction, however, we see a $C(0)=0$ region and a $C(0)=-2$ region, showing the \{001\} phase has been stabilized, as expected, by such octahedral rotations, through introduction of $xy \leftrightarrow yz$ and $xy \leftrightarrow xz$ inter-orbital W-W hopping. Thus, although octahedral rotations about the $\hat{b}$-axis stabilize the topology of the lowest band, they have also created a $C(0)=0$ phase for alignment of the ferromagnetically-aligned Cr moments in the \{111\} crystallographic direction. In the absence of such rotations, there is a $C(0)=\pm4$ phase for this magnetic order and orientation even in the absence of an $xz/yz$ orbital shift. This indicates these octahedral rotations are, in fact, detrimental to the exotic $C(0)$ band topology shown in Fig. \ref{fig5}.   

\section{Discussion and Conclusion}

We have found that a tight-binding model for t$_{2g}$ orbitals on a square lattice -- relevant to monolayers of a class of double perovskites including Ba$_2$FeReO$_6$, Sr$_2$FeMoO$_6$, Sr$_2$CrMoO$_6$, and Sr$_2$CrWO$_6$ -- can be topologically unstable. Specifically, using Sr$_2$CrWO$_6$ as an example, we computed the ground state energy when classical moments are aligned ferromagnetically in each of the five distinct high-symmetry directions, as well as the Chern number for the lowest band of the dispersion, to compute a phase diagram as a function of tetragonal distortion $\chi_{tet}$ and inter-orbital hopping $t_m$. For sufficiently large tetragonal compression in the $\hat{z}$ direction, the classical moments prefer to align in the \{001\} crystallographic direction.  At weaker tetragonal compression, the moments prefer to align in the \{100\} and \{110\} crystallographic directions.  When the moments prefer to align in the \{001\} direction, the dispersion exhibits line nodes that result in topological instability of this magnetic phase.  These line nodes can be understood with an effective model as resulting from a lack of interorbital hopping between $xy$ and $yz$/$xz$ orbitals.  When the moments prefer to align in the \{100\} crystallographic direction, the dispersion exhibits point nodes. An effective 2-band model for the \{100\} phase also exhibits point nodes.  Finally, when the moments prefer to align in the \{110\} crystallographic direction, the 9-band model exhibits point nodes, but an effective model for this phase suggests the system should actually exhibit line nodes.  This suggests physics not captured by the effective model leads to some symmetry-breaking terms that result in point nodes rather than line nodes in the 9-band model, and this will be explored in future work. 

We have also found, using a parameter set relevant to Sr$_2$CrWO$_6$ as our example, that topological phases can be stabilized in the 9-band tight-binding model with physically-relevant, symmetry-breaking terms to yield Chern numbers for the lowest band that are greater in magnitude than one. Adding a symmetry-allowed shift of $xz$/$yz$ orbitals up out of the monolayer plane allows Chern numbers of $\pm2$, $\pm4$, $\pm6$, and $\pm8$ to be reached for the lowest band of the model dispersion through a combination of sufficient tetragonal compression of W octahedral cages and application of weak in-plane magnetic fields. Effective models are shown to capture some of this topological physics, such as a relative change in Chern number in the \{001\} phase of 4 with increasing tetragonal compression. Even without the additional $xz$/$yz$ orbital shift, pushing the system into a ground state where Cr moments are ferromagnetically-aligned in the $\{001\}$ crystallographic direction via sufficiently strong tetragonal compression of W octahedra, and then applying in-plane magnetic fields to tip the Cr moments away from the $\hat{z}$ axis can realize a large region in parameter space corresponding to Chern number of $\pm4$ for the lowest band of the dispersion. These results suggest that the model exhibits exotic Chern numbers somewhat generically. 

Small staggered octahedral rotations about the $\hat{z}$ axis stabilize the $\{110\}$ phase shown in Fig. \ref{fig2} and this can be understood in part through the effective two-band model for this phase. Without the rotations, coefficients for $|xz\rangle$ and $|yz\rangle$ states in expressions for $|J_{110}=-3/2\rangle$ and $|J_{110}=-1/2\rangle$ balance one another to cancel out off-diagonal terms. The rotations introduce imbalances in hoppings that prevent this cancellation, leading to point nodes in the effective model and removal of nodes in the full model. In contrast, staggered octahedral rotations about the $\hat{z}$-axis do not stabilize the $\{001\}$ phase, since such octahedral rotations do not introduce the $xy$ to $xz$/$yz$ inter-orbital hopping the effective two-band $\{001\}$ model shows is needed to stabilize this phase. Although these staggered octahedral rotations do not appear to generate any additional exotic topological phases, the $C(0)=\pm4$ phase shown in Fig. \ref{fig5} should be largely-unaffected as it occurs over a wide range in $\theta$ and $\phi$. In contrast, the $C(0)=\pm4$ phase appears to be sensitive to staggered octahedral rotations about the $\hat{b}$-axis, which leads to a $\{111\}$ ground state magnetic order with $C(0)=0$ instead of $C(0)=\pm4$, suggesting this phase is no longer present or significantly reduced. This kind of staggered octahedral rotation does stabilize a $C(0)=\pm2$ phase for $\{001\}$-aligned ferromagnetic ordering of Cr moments for weak tetragonal distortion, however, which is interesting but possibly difficult to access in experiments.
 
Further exploration is warranted to identify the upper bounds on Chern number values in this model and gain deeper understanding of the origins of this exotic topology. Past work on double perovskite bilayers\cite{cook000} indicates higher Chern numbers emerge naturally from relative phases introduced through hopping between different states of the $J={3\over 2}$ manifold. This earlier work and the results presented here may indicate that states of higher $J$ manifolds can more generally lead to higher Chern numbers in combination with certain lattice geometries. Further work is required to explore this hypothesis and possibly make these relationships concrete. Later work will also include investigation of these topological phases when correlations are present, identification of other approaches to realizing exotic topological phases in this model and other models for TMOs, and investigation of various approaches to opening gaps in these systems to realize quantum anomalous Hall insulator phases.

It should be acknowledged that realization of ordered double perovskite monolayers grown in the \{001\} crystallographic direction is, at present, a daunting task. It would require laying down a checkerboard of B and B' ions, which might be prone to severe anti-site disorder despite ionic size mismatches due to B being a 3d ion and B' being a 4d or 5d ion. This work therefore serves at present to show the potential promise of TMOs with valence electrons in manifolds of higher angular momentum $J$ for realization of higher Chern number topology for individual bands. As fabrication of ordered DP monolayers in the \{001\} crystallographic direction becomes feasible, however, further investigation of physics discussed here and related issues relevant to experimental study will be warranted. 

\begin{acknowledgements}
This work was supported by the Natural Sciences and Engineering Research Council of Canada. AMC would also like to acknowledge helpful discussions with Prof. Arun Paramekanti and Dr. Vijay Shankar Venkataraman during work on this manuscript.
\end{acknowledgements}

\bibliography{scwo61215}

\end{document}